# Multimodal Integrated Knowledge Transfer to Large Language Models through Preference Optimization with Biomedical Applications

Zhanliang Wang[1,2#], Da Wu[1#], Quan Nguyen[1,3], Zhuoran Xu[1,4], Kai Wang[1,5*]

[1] Raymond G. Perelman Center for Cellular and Molecular Therapeutics, Children's Hospital of Philadelphia, Philadelphia, PA 19104, USA

[2] Applied Mathematics and Computational Science Graduate Program, University of Pennsylvania, Philadelphia, PA, 19104, USA

[3] Bioengineering Graduate Program, University of Pennsylvania, Philadelphia, PA 19104, USA

[4] Genomics and Computational Biology Graduate Program, University of Pennsylvania, Philadelphia, PA 19104, USA

[5] Department of Pathology and Laboratory Medicine, Perelman School of Medicine, University of Pennsylvania, Philadelphia, PA 19104, USA

*: correspondence should be addressed to wangk@chop.edu.

#: equal contribution.

## ABSTRACT

The scarcity of high-quality multimodal biomedical data limits the ability to effectively fine-tune pretrained Large Language Models (LLMs) for specialized biomedical tasks. To address this challenge, we introduce MINT (Multimodal Integrated kNowledge Transfer), a framework that aligns unimodal large decoder models with domain-specific decision patterns from high-quality multimodal biomedical data through preference optimization. While MINT supports different optimization techniques, we primarily implement it with the Odds Ratio Preference Optimization (ORPO) framework as its backbone. This strategy enables the aligned LLMs to perform predictive tasks using text-only or image-only inputs while retaining knowledge learnt from multimodal data. MINT leverages an upstream multimodal machine learning (MML) model trained on high-quality multimodal data to transfer domain-specific insights to downstream text-only or image-only LLMs. We demonstrate MINT's effectiveness through two key applications: (1) Rare genetic disease prediction from texts, where MINT uses a multimodal encoder model, trained on facial photos and clinical notes, to generate a preference dataset for aligning a lightweight decoder-based text-only LLM (Llama 3.2-3B-Instruct). Despite relying on text input only, the MINT-derived model outperforms models trained with Supervised Fine-Tuning (SFT), Retrieval-Augmented Generation (RAG), or direct preference optimization (DPO), and even outperforms much larger foundation model (Llama 3.1-405B-Instruct). (2) Tissue type classification using cell nucleus images, where MINT uses a vision-language foundation model as the preference generator, containing knowledge learnt from both text and histopathological images to align downstream image-only models. The resulting MINT-derived model significantly improves the performance of Llama 3.2-Vision-11B-Instruct on tissue type classification. In summary, MINT provides an effective strategy to align unimodal LLMs with high-quality multimodal expertise through preference optimization. Our study also highlights a hybrid strategy that grafts the strength of encoder models in classification tasks into large decoder models to enhance reasoning, improve predictive tasks and reduce hallucination in biomedical applications.

## INTRODUCTION

Language Models (LLMs) have achieved remarkable advances in natural language understanding and reasoning, powered by the scaling properties of the Transformer architecture[1,2] and pretraining on large-scale, general-purpose text corpora[3-5]. Compared to earlier natural language processing (NLP) models, such as those from the Bidirectional Encoder Representations from Transformers (BERT)[6], LLMs exhibit superior linguistic capabilities, more coherent long-range dependencies, context handling[6-9] and improved generalization across diverse tasks. Recent models such as Llama 3.1[7]/3.2[8]/3.3[9] and DeepSeek V3[10]/R1[11] extend these capabilities further with context windows of up to 128K tokens, enabling more accurate processing of lengthy, detailed, and often noisy biomedical documents. This progress has expanded their potential for a broad range of biomedical applications, including summarizing noisy clinical documents[12-15], managing patient scheduling[16-18], and addressing highly specialized tasks such as rare disease prediction and gene prioritization[19,20]. Despite these advances, adapting LLMs to domain-specific biomedical tasks remains challenging, especially when multimodal data are scarce, supervision is sparse, or the task requires specialized clinical reasoning.

Supervised fine-tuning (SFT)[21,22] is the most commonly employed approach for adapting LLMs to domain-specific tasks using task-specific datasets that are typically much smaller than those used during pretraining. SFT has proven effective for linguistically driven applications[23-25]. For example, Yang et al. developed PhenoGPT[26] to derive phenotypes from long and complex clinical notes by fine-tuning LLMs on data annotated with Human Phenotype Ontology (HPO) terms, achieving improved performance over the foundation LLMs. While such applications demonstrate benefits of SFT in biomedical NLP tasks, the approach faces notable limitations when applied to more complex tasks that require structured prediction and complex logical reasoning, such as rare disease diagnosis or gene prioritization[27-29]. These tasks often require open-ended generation of precise, long-tail clinical terminology and reasoning beyond the label coverage provided by available supervision, which can be especially challenging for auto-regressive LLMs that generates responses token by token[2]. In principle, some settings could be reduced to classification given an exhaustive and well-curated term inventory; however, obtaining sufficiently comprehensive, high-quality annotations at scale is often infeasible in biomedical research, where expert labeling is costly and many disease or tissue categories are rare. The challenge is amplified when labeled data are scarce, as is often the case in biomedical domains. Moreover, indiscriminate fine-tuning on medical datasets can have unintended consequences, potentially weakening the model's inherent linguistic and logical reasoning abilities significantly[30-32]. These limitations becomes even more pronounced in zero-shot or few-shot scenarios, where models must handle previously unseen disease classes or tissue types, which are increasingly encountered in real-world clinical practice[33].

To tackle these challenges, we propose Multimodal Integrated Knowledge Transfer (MINT), a framework that facilitates aligning LLMs with domain-specific patterns from multimodal biomedical databases (e.g., images, audios, videos, clinical texts) while preserving their original linguistic and reasoning capabilities.

MINT can be implemented with different preference optimization techniques including Direct Preference Optimization (DPO)[34] and Odds Ratio Preference Optimization (ORPO)[35]. Preference optimization aligns a language model by learning from pairwise preferences (a chosen response versus a rejected response) rather than relying solely on token-level supervision. In the RLHF paradigm, this is often implemented by training a reward model and optimizing the policy with reinforcement learning (e.g., PPO), whereas recent methods such as DPO and ORPO optimize the model directly from preference pairs without requiring an explicit reward model. In this paper, we primarily use MINT with ORPO as our default backbone, which we refer to simply as "MINT" hereafter for brevity. The MINT framework comprises two key components: upstream preference learning dataset construction and downstream LLM model training. In the upstream pipeline, a multimodal ML model is trained to generate a preference learning dataset, including a list of preferred responses and a list of unfavored responses. In the downstream pipeline, these preference-labeled inputs are used for aligning LLM using ORPO to prioritize preferred responses over unfavored ones while preserving its general linguistic and reasoning capabilities.

To showcase its power, we demonstrate that the MINT framework significantly enhances LLM performance in two important biomedical tasks: phenotype-driven rare disease prediction and image-based tissue type classification. For the first task, we utilize GestaltMML[36], a multimodal Transformer model trained on facial images, clinical texts, and demographic information, to construct a preference learning dataset. Our results show that the MINT-aligned model outperforms widely used methods such as SFT and Retrieval-Augmented Generation (RAG)[37-39] in the task of performing text -only rare disease prediction, and even demonstrates strong performance on previously unseen disease classes in zero-shot settings. The second application of MINT focuses on tissue type classification from cell nucleus images using the PanNuke Database[40,41] and Pathology Language-Image Pretraining (PLIP)[42] model as the upstream multimodal preference dataset generator. By aligning Llama 3.2-Vision-11B-Instruct with preferences generated from PLIP, MINT outperforms other widely used techniques for image-based tissue type classification, such as Supervised Fine-Tuning (SFT). Together, we believe that the MINT framework can be readily extended to other biomedical applications, enabling the efficient transfer of knowledge from domain-specific multimodal datasets to LLMs or other large decoder models. This approach serves as a similar but alternative technique to RAG, which combines the strength of encoder in classification tasks with large decoder models.

## RESULTS

### *Overview*

In MINT, each task utilizes an upstream multimodal machine learning (MML) model to generate a preference learning dataset, which consists of most-likely and least-likely labels for each sample. The downstream large language model (LLM) then serves as the target for multimodal knowledge transfer (**Figure 1)**. Throughout the Results section, we refer to our default implementation (MINT with ORPO) simply as 'MINT' for brevity, while the alternative implementation with DPO is referred to as "DPO".

To demonstrate the effectiveness of the MINT framework, we evaluate its performance on two essential tasks: rare genetic disease prediction and tissue type classification. A summary of datasets, upstream multimodal models and downstream LLMs used in this study is given in **Table 1**. For rare disease prediction, the upstream model is GestaltMML, a Vision-and-Language Transformer-based multimodal ML model, while the downstream model is the text-only Llama-3.2-3B-Instruct since our input will be clinical texts only. For tissue type classification, the upstream model is Pathology Language-Image Pretraining (PLIP), a vision-language foundation model designed for pathology AI, whereas the downstream model is Llama 3.2-Vision-11B-Instruct, whose input will be images with a trivial question such as "What is shown in Figure?"

Phenotype-driven disease prediction[43-46] is a task that makes predictions of rare diseases based on patients' clinical phenotypes, and such predictions can assist physicians in selecting diagnostic modalities or to facilitate clinical labs in finding causal genes/mutations[47,48]. Such clinical phenotypes can include facial photos, clinical notes, lab tests, and other imaging modalities. Several machine learning models have been developed to utilize facial photos for disease prediction, such as DeepGestalt[49] and GestaltMatcher[50], which are based on facial image analysis, as well as GestaltMML[36], a multimodal model that integrates facial images, demographic information, and clinical phenotypes such as Human Phenotype Ontology (HPO) terms[47].

GestaltMML is a Transformer-based multimodal machine learning (ML) model that combines the ViLT (Vision-and-Language Transformer[51]) with the GestaltMatcher Database (GMDB)[50,52]. Unlike earlier multimodal models[50] that use distinct image and text models with simple fusion, GestaltMML employs a unified Transformer encoder[2] to model interactions between facial images and structured clinical texts.

PanNuke is a cell nuclei instance segmentation and classification dataset with exhaustive nuclei labels across 19 different tissue types. The dataset consists of 481 visual fields sampled from more than 20K whole slide images at different magnifications, with a total of 205,343 labeled nuclei, each with an instance segmentation mask. PLIP, built upon the widely recognized Contrastive Language-Image Pre-training (CLIP) model[53], is the first vision-and-language foundation model designed for pathology AI.

Our goal here is not nuclei segmentation but predicting the tissue type (such as "bile duct" and "colon") from cell nucleus images. We show that MINT, when applied to Llama 3.2-Vision-11B-Instruct, outperforms other widely used techniques for image-based tissue type classification, such as Supervised Fine-Tuning (SFT).

We assess MINT's performance by comparing it against the foundation LLM and other approaches, including supervised fine-tuning (SFT), retrieval-augmented generation (RAG), and direct preference optimization (DPO) on four key evaluation metrics: Hallucination Free Accuracy (HFA), Top-N (N= 5 or 10) Accuracy, Top-1 Accuracy, and Coverage-Avoidance Ratio (CAR). Hallucination-Free Accuracy measures the rate at which a language model does not produce hallucinations, with 100% indicating no hallucinations at all. For "Hallucination", we mean the model either does not follow the instruction at all or produces fabricated labels. The CAR measures the model's capability to distinguish a list of preferred responses over the least-likely responses. Note that RAG is implemented and evaluated for the text-based rare disease prediction task only in our study.

***MINT Enhances Rare Disease Prediction Accuracy by LLM***

We first evaluate it on a rare disease prediction task using text-based clinical phenotypes. In this task, the downstream LLM is given a short clinical summary (a few hundred words) that includes demographic information such as the patient's age, gender, and ethnicity, along with observed phenotype features. For instance, a typical input might describe a 1-year-old female patient of Caucasian descent presenting with developmental delay, overgrowth, scoliosis and cardiovascular abnormalities. Based on this input, the model generates a ranked list of ten possible rare diseases that best match the provided features. The output is structured in a consistent format, with each entry containing the full disease name and, when applicable, a commonly used abbreviation. Examples of generated predictions include conditions such as "Phelan-McDermid syndrome (PHMDS)" and "Williams-Beuren syndrome (WBS)." This structured response helps clinicians quickly interpret model outputs and assess diagnostic relevance.

To benchmark prediction performance, we compared MINT against the base model without any domain knowledge enhancement, as well as several existing model enhancement techniques, including Retrieval-Augmented Generation (RAG), Supervised Fine-Tuning (SFT), and SFT followed by Direct Preference Optimization (DPO), which we refer to as DPO hereafter for simplicity. To further validate MINT's capabilities against domain-specific baselines, we additionally included a comparison with MedGemma-1.5-4b-it[54], a specialized medical LLM. **Figure 2** presents the evaluation results of our best-performing configuration, using the Llama 3.2-3B-Instruct enhanced with MINT under a balanced preference setting (AoR = 1.0), i.e., 10 accepted vs. 10 rejected responses. The AoR is defined as the ratio of accepted to rejected responses when building the preference-learning datasets (see the Methods section for details). For this evaluation, we utilize a dataset of 6522 training samples and 386 testing samples compiled from the GestaltMatcher Database (GMDB). The choice of how many responses to accept and reject was determined through considerations of the total number of disease labels (~528). The figure summarizes diagnostic accuracy, robustness under different preference ratios, and general language understanding capability.

We compare various enhancement strategies applied to base Llama 3.2-3B-Instruct model, using our four key metrics: Hallucination-Free Accuracy (HFA), Top-10 accuracy, Top-1 accuracy, and Coverage-Avoidance Rate (CAR) (**Figure 2a**). Our proposed method, MINT, consistently achieves the best performance across all metrics. Notably, MINT raises Top-10 accuracy from 5.19% (base model) to 52.99%, outperforming other methods such as RAG (6.52%), SFT (37.53%) and DPO (38.49%). . Furthermore, as shown in **Table S7,** MINT significantly surpasses the specialized model MedGemma-1.5-4b-it (32.45%), demonstrating that preference alignment can yield superior diagnostic performance compared to domain-specific pre-training, even with fewer parameters (3B vs 4B). The Coverage-Avoidance Rate (CAR) metric further highlights MINT's strength. CAR is a harmonic mean between Top-k coverage and Bottom-q avoidance, assessing whether the model can prioritize likely diseases while avoiding implausible ones. On this metric, MINT achieves a CAR of 0.2877, significantly higher than both SFT (0.1665) and DPO (0.2808). This indicates that MINT not only improves recall of true conditions in the top ranks, but also minimizes the inclusion of clinically irrelevant diseases, thereby improving reliability for real-world diagnostic support.

We also examine the robustness of MINT under varying Acceptance-over-Rejection (AoR) ratios, which reflect the balance between accepted and rejected samples in the preference optimization phase (**Figure 2b**). We observe that performance improves as the AoR approaches a balanced ratio (10 accepted vs. 10 rejected). This trend suggests that a well-balanced preference dataset provides clearer and more stable gradient signals for learning disease ranking, while overly imbalanced settings such as too many rejected candidates, may dilute the optimization signal and reduce learning efficiency. These results offer practical guidance for designing preference-based supervision strategies in downstream medical applications.

To address the possibility that MINT-enhanced models may lose the general language understanding capabilities, we evaluate it on the H6-benchmark , a comprehensive evaluation suite comprising six-widely-used datasets: MMLU[55], TruthfulQA[56], HellaSwag[57], Winogrande, ARC[58] and GSM8k[59]. In total, the H6-benchmark contains approximately 113,000 diverse samples that test various aspects of language understanding and reasoning through multiple-choice questions. The results demonstrate that MINT does not compromise the base model's generalization ability. Performance is maintained or modestly changed across all benchmarks, suggesting that MINT's improvements in rare disease prediction are not at the expense of broad language or reasoning skills. This balance is especially important for models expected to operate in open-domain or multi-task settings.

To validate the generality of MINT across model sizes and architectures, we report full results in **Table S2**, which includes experiments on Llama 3.2-1B, 3B, 8B, Gemma 2-2B, 9B, and Llama 3.1-8B,70B, 405B. Due to computational constraints, only inference is performed on the largest models (70B and 405B). Across all scales and families, MINT consistently improves both the Top-10 and Top-1 accuracy. For example, on Llama 3.2-1B-Instruct, Top-10 accuracy increases from 4.24% to 34.28%, yet on Gemma-2-9B, Top-10 accuracy improves from 12.37% to 33.80%. Similarly, we observed a large increase of CAR values under different MINT model sizes and architectures, outperforming all the corresponding baselines. These gains are consistently accompanied by near-perfect HFA values, often exceeding 99%, indicating that MINT substantially enhances accuracy while maintaining high factual consistency and avoiding hallucinated outputs. This is to be expected since the upstream model GestaltMML is an encoder model that puts strong constraints on the prediction to ~520 disease labels.

### ***Additional Validation and Case Study for Rare Disease Prediction***

Our results presented in the previous section are performed by splitting the original GMDB datasets into training and testing subsets. We recognize that cross-validation can be susceptible to overfitting, especially if the data used for validation is not representative of the true distribution of data the model will encounter in real-world scenarios. Indeed, the clinical phenotypes from the GMDB database often comprise of a few phenotype terms in the form of HPO, rather than natural texts that are written by human experts. To assess the robustness and generalizability of MINT in real-world clinical contexts, we conducted external validation using Phenopacket-derived clinical notes[60]. This dataset consists of 5980 high-quality synthesized clinical notes that simulate realistic clinical narratives by converting standardized phenotype profiles ("Phenopackets") into natural language. This introduces the linguistic variability and contextual noise commonly observed in clinical practice. We divided this dataset into two subsets: one containing set of diseases that overlap with the GMDB training set (1638 samples with 72 overlapping diseases) and another that are non-overlapping (disjoint) from the training set (4342 samples with 396 disjoint diseases). This division allows us to assess both in-distribution performance and zero-shot generalization to unseen disease classes.

As shown in **Table 2**, MINT demonstrates strong performance across both subsets. For the overlapping disease subset, MINT substantially outperforms all baselines, achieving 66.91% Top-10 accuracy and 47.56% Top-1 accuracy, compared to the base model's 5.13% and 3.42%, respectively. The performance gap is particularly notable compared to other enhancement techniques: DPO (60.99% Top-10, 26.56% Top-1), SFT (46.40% Top-10, 14.65% Top-1), and RAG (30.11% Top-10, 26.45% Top-1).

For the more challenging disjoint disease subset, which evaluates zero-shot generalization to completely unseen disease classes, performance decreases across all methods as expected. MINT achieves 10.48% Top-10 accuracy and 7.00% Top-1 accuracy. Interestingly, in this zero-shot scenario, RAG shows competitive performance with 24.17% Top-10 accuracy and 13.80% Top-1 accuracy, highlighting the complementary strengths of retrieval-based approaches when dealing with previously unseen diseases. This suggests that combining MINT with retrieval mechanisms could yield further improvements for zero-shot prediction scenarios. All methods maintain near-perfect Hallucination-Free Accuracy (HFA) rates above 99%, indicating strong factual consistency across approaches.

These extensive validation experiments demonstrate that MINT improves rare disease prediction across diverse text sources when diseases overlap with the training distribution. For zero-shot scenarios with previously unseen diseases, retrieval augmented approaches offer complementary strengths, suggesting that combining preference optimization with retrieval mechanisms could be particularly effective for comprehensive rare disease prediction systems.

To better understand the performance drop on the disjoint split, we analyzed the disease-label coverage in the constructed preference dataset (**Table S5**). Among the 396 unique disjoint ground-truth diseases, 377 (95.2%) never appear in either the preferred (chosen) or unfavored (rejected) candidate lists produced in our upstream preference construction, and only 19 (4.8%) appear in both lists. This indicates that, for fully disjoint diseases, preference-based alignment provides little direct supervision signal for the target labels and primarily reinforces the model toward the disease space covered by the preference candidates. As a result, preference optimization (and SFT) can underperform the base model on fully unseen disease categories, while retrieval-based methods (RAG) remain more robust by injecting disease-specific descriptions at inference time, suggesting that preference alignment and retrieval are complementary for real-world rare disease prediction. These results suggest that MINT is not intended as a standalone solution for open-world diagnosis under fully unseen disease categories; instead, it is best viewed as complementary to retrieval-based mechanisms in such settings.

***MINT Substantially Enhances Tissue Type Classification Accuracy by LLM***

We further assess the capability of MINT on a vision-centric biomedical classification task: tissue type prediction using the PanNuke dataset. The objective is to identify the most likely tissue of origin given a histological image, selected from a fixed set of 19 tissue types. This task is cast into a vision-language interface by providing each model with two inputs: (i) an image patch extracted from a nucleus-level H&E-stained slide, and (ii) a standardized prompt that asks, *"Based on the morphological features observed in the provided nucleus image, which tissue types could this nucleus originate from?"* The model is instructed to generate a ranked list of exactly five candidate tissue types. While the prompt follows the format of a typical Visual Question Answering (VQA) task, it is semantically constant across all samples and thus does not provide additional context. As a result, the task relies entirely on visual modality from decision-making, making it a faithful benchmark of visual reasoning capacity.

We summarize the results of applying MINT to the downstream Llama 3.2–11B-Vision-Instruct model and three common enhancement baselines: SFT, DPO, and the base model (**Figure 3a**). For this evaluation, we utilize 5436 training images and 2330 testing images from the PanNuke dataset. Full numerical results with standard deviations across three random seeds are provided in **Table S3**. The base vision-language model achieves moderate baseline performance, with a Top-5 accuracy of 32.21% and Top-1 accuracy of 16.96%, likely reflecting pre-existing visual understanding from its large-scale pretraining. Upon applying MINT, we observe a substantial performance gain where Top-5 accuracy rises to 57.58% and Top-1 accuracy improves to 28.41%, representing a nearly doubling of accuracy compared to the base model. In addition to accuracy gains, MINT achieves the highest Coverage-Avoidance Rate (CAR) of 0.5203, slightly surpassing the best baseline (DPO, 0.5196) and substantially exceeding the base model (0.2965). The CAR is computed using 5 accepted responses and 5 rejected responses due to the small label set (19 classes), reflects the model's ability to both surface relevant tissue types in the top predictions and avoid implausible categories in the bottom ranks.

Other techniques, including SFT, DPO, show moderate gains; For example, both SFT and DPO improve Top-5 accuracy to 41.16%, yet fall short of MINT across all key metrics. These results reinforce MINT's strength in enhancing multimodal models across both text-based and image-based biomedical tasks. Note that due to the small size of the label space, we do not explore varying Acceptance-over-Rejection (AoR) ratios in this task and instead fix the preference optimization setting to a balanced 5 accepted vs. 5 rejected labels.

To examine whether MINT and other fine-tuning strategies compromise the general capabilities of vision-language models (VLMs), we further evaluate their performance on SEED-Bench[61], a large-scale benchmark consists of 19,000 multiple choice questions with high-quality human annotations SEED-Bench spans ten core evaluation dimensions, including instance interaction, instance identity, instance attributes, instance location, instance counting, scene understanding, spatial relation, text understanding, and visual reasoning. As shown in **Figure 3b**, the performance of all fine-tuned models—MINT, SFT, and DPO—closely aligns with that of the base model across all axes. These findings indicate that none of the fine-tuning approaches, including MINT, degrade the general-purpose visual-language reasoning ability of the foundation model, while MINT simultaneously delivers substantial performance improvements on the domain-specific task of tissue type classification.

***Case Studies: MINT Resolves Ambiguity in Challenging Diagnostic Scenarios***

*Differentiating Histological Similarity (Vision Task)*

We present detailed case studies demonstrating MINT's effectiveness in distinguishing between Colon[62] and Bile Duct[63-65] tissue images (**Figure 4**), which represent a particularly challenging discrimination task due to their histological similarities. Both tissues exhibit comparable epithelia structures and are functionally related within the digestive system[66,67]. Colon epithelium consists of columnar cells with microvilli that enhance nutrient absorption and water reabsorption, while bile ducts are lined by cholangiocytes with similar microvilli that aid bile modification and transport[68,69].

We first focused on the distinction between Colon and Bile Duct tissue images, a particularly challenging task due to their histological similarities. Both tissues exhibit comparable epithelial structures and are functionally related within the digestive system. As shown in **Figure 4,** the SFT baseline showed considerable confusion, frequently ranking the incorrect tissue higher than the ground truth (e.g., assigning an average rank of 2.00 to the confuser class Colon when the ground truth was Bile Duct).

In contrast, MINT consistently assigned Rank 1 to the correct tissue type across test samples while pushing the visually similar confuser class to substantially lower ranks (Average Rank 5.75 for Colon confusers). This discriminative capability is further detailed in Supplementary **Figure S1B**, which illustrates a specific instance where MINT suppressed the Bile Duct prediction for a Colon sample. By explicitly incorporating both chosen (positive) and rejected (negative) tissue types during training, MINT learns subtle morphological features that differentiate visually similar tissues, whereas SFT learns solely from positive samples without explicitly accounting for confusers.

*Resolving Phenotypic Ambiguity (Text Task)*

We extended our analysis to the text domain to examine how MINT handles phenotypic ambiguity in rare disease prediction. Specifically, we focused on the distinction between Cornelia de Lange Syndrome (CdLS)[70] and phenotypically similar conditions such as Hypertrichosis[71]. This comparison represents a particularly challenging diagnostic task for text-only models[60] due to the significant overlap in generic systemic phenotypes. Both conditions frequently present with hirsutism (excessive hair growth), hypertrichosis of the eyebrow, and varying degrees of developmental delay[72]. However, CdLS is distinctively characterized by a specific facial gestalt, which are the features often referred to as "classic" or pathognomonic including synophrys (joined eyebrows), long philtrum, thin upper vermilion, and anteverted nares[36,52].

To illustrate MINT's superior discriminative capability, we selected representative test samples describing patients with CdLS who presented with a mix of generic symptoms (e.g., "Intellectual disability", "Hirsutism") and specific facial features ("Synophrys"). As illustrated in **Figure S1A**, the SFT baseline was misled by the high frequency of generic terms. It consistently prioritized Hypertrichosis (Rank 1), likely because the token "hirsutism" in the input strongly correlates with the disease label "Hypertrichosis" in general medical corpora, overpowering the more specific but less frequent facial descriptors[73]. Consequently, SFT assigned the correct diagnosis (CdLS) to a lower rank (Rank 2), failing to resolve the ambiguity.

In contrast, MINT successfully resolved this phenotypic conflict. It correctly identified Cornelia de Lange Syndrome as Rank 1 and suppressed the confuser class Hypertrichosis. This improvement stems directly from the knowledge transfer via preference optimization. The upstream multimodal model (GestaltMML[36]), having been trained on facial images where the "CdLS gestalt" is visually distinct from simple Hypertrichosis[52,74], generated preference pairs that explicitly "rejected" Hypertrichosis for patients with these specific facial features. MINT effectively learned to assign higher attention weights to discriminative facial terms like "Synophrys" and "Long philtrum" over generic terms like "Hirsutism." By mimicking the visual reasoning of the upstream expert, MINT enables the LLM to reject plausible but incorrect "confuser" diseases that would otherwise mislead a standard text-based model.

*Sensitivity Analysis: Impact of Preference Dataset Size across Tasks*

To evaluate the data efficiency of the MINT framework and determine the volume of multimodal knowledge required for effective alignment, we performed a sensitivity analysis on both biomedical tasks. While keeping the upstream multimodal models fixed, we trained the downstream models using random subsets of the preference pairs, ranging from 20% to 100% of the total training data.

As shown in **Table S6,** we observed a consistent scaling trend across both text-only and vision-language tasks. For Rare Disease Prediction, downstream performance improved significantly with data volume.

With only 20% of the data (~1,304 samples), the model achieved 19.32% Top-10 accuracy, outperforming the base model (5.19%). Notably, at 40% data usage, MINT reached 34.60% Top-10 accuracy, approaching the performance of standard SFT trained on the full dataset (37.53%).

Similarly, for Tissue Type Classification, MINT demonstrated strong efficiency. Training with just 20% of the preference data (~1,087 samples) improved Top-5 accuracy to 38.45% (compared to the Base Model's 32.21%). The performance steadily increased to 57.58% with the full dataset. These results indicate that the contrastive signals provided by the preference optimization are highly efficient at transferring domain knowledge across modalities. While MINT yields substantial gains with limited data, utilizing the full breadth of the preference dataset remains essential for maximizing predictive accuracy in complex, high-dimensional biomedical tasks.

## METHODS

### *Evaluation Metric*

In this paper, we assess the capabilities of large language models (LLMs) on rare genetic diseases prediction using text description of clinical phenotypes across three main outcomes: (HFA: Hallucination-Free Accuracy) whether the LLM can produce a meaningful response to a query, specifically a list of real disease names; (Top-N) if (HFA) is successful, whether the LLM's response of N choices includes the correct disease prediction (true disease name); (Top-1) if (HFA) is successful, whether the LLM can accurately select the correct disease prediction. These tasks follow a logical progression: outcomes (Top-N) and (Top-1) are only applicable if (HFA) is achieved, meaning if (HFA) fails, (Top-N) and (Top-1) will also fail. For rare disease prediction, we choose N = 10, while for tissue type classification, we chose N = 5.

Motivated by the preference optimization framework (DPO and ORPO), we also evaluate the model's ability to prioritize likely diseases in the Top-$k$ preferred responses while avoiding Bottom-$q$ unlikely responses. Let $R_i$ denote the $i$th response by LLM, consisting of a list of rare disease names. Let $T_{k,i}$ denote the top-$k$ disease names ($k$ labels with highest probabilities) predicted by upstream Multimodal ML model. If the true label if not at the first position, then we will swap the true label with the one first position. Let $B_{q,\ i}$ denote the bottom-$q$ disease names predicted by upstream Multimodal ML model. Similarly, it consists of diseases with $q$ lowest probabilities among all the diseases.

Next, we define the Top-$k$ coverage rate and Bottom-$q$ avoidance rate, denoted by $C_{k,i}$ and $A_{q,i}$ respectively, by

$$C_{k,i} = \frac{|T_{k,i} \cap R_i|}{|R_i|} \qquad \text{and} \qquad A_{q,i} = 1 - \frac{|R_i \cap B_{q,i}|}{|R_i|},$$

respectively. Finally, we define the overall Coverage-Avoidance Rate (CAR) to be the weighted harmonic mean between Top-$k$ coverage rate and Bottom-$q$ avoidance rate. Mathematically, for $\lambda \geq 0$ ,

$$CAR_{(k,q,\lambda)} = \frac{1}{N}\sum_{i=1}^{N}\frac{(1+\lambda)C_{k,i}\cdot A_{q,i}}{\lambda C_{k,i}+A_{q,i}}$$

The weighting parameter $\lambda$ can be adjusted further to balance the importance between Top-$k$ coverage and Bottom-$q$ avoidance rate. Note that when $\lambda = 1$, both $C_{k,i}$ and $A_{q,i}$ are equally weighted, and CAR is just the naive harmonic mean. When $k \leq q$, we can set $\lambda > 1$ to give $C_{k,i}$ more weight. Likewise, when $k > q$, we can set $\lambda < 1$ to give $A_{q,i}$ more weight. We also denote $k/q$ by Acceptance-over-Rejection (AoR) ratio**.**

Finally, we benchmark all the add-on techniques using six widely recognized datasets, including ARC (Abstract Science Reasoning)[58], HellaSwag (Commonsense Inference)[57], MMLU (Massive Multitask Language Understanding)[55], TruthfulQA (Detecting False Information)[56], Winogrande (Context-Based Inference)[75], and GSM8K (Mathematical Reasoning)[59]. ARC (AI2 Reasoning Challenge) assesses scientific reasoning with grade-school science questions. HellaSwag tests commonsense inference through narrative continuation tasks. MMLU (Massive Multitask Language Understanding) evaluates multitask knowledge and reasoning across 57 diverse subjects. TruthfulQA measures the model's ability to avoid reproducing falsehoods. Winogrande focuses on common-sense reasoning with ambiguous pronoun resolution tasks. GSM8K asks the model with multi-step grade school math problems. All evaluations use a 5-shot setting to ensure consistency and rigor, which verifies that the techniques retain the original language understanding and reasoning capabilities of the base model after fine-tuning. This evaluation aims to determine whether the add-on technique retains the original language and reasoning capabilities of the base model.

***Automated Evaluation Program***

Given the extensive number of experiments involved, we develop an automated evaluation framework, LLM-Eval, to systematically and rigorously assess the responses generated by LLMs. The evaluation process comprises multiple steps to ensure the plausibility, relevance, and ranking quality of the model's diagnoses. It begins with HFA, which checks whether the model generates plausible disease names by filtering out extreme hallucinations, such as nonsensical phrases or invalid outputs. Once HFA is passed, Top-N evaluates whether the ground truth disease is included in the ranked list of $N$ possible diseases generated by the model, while Top-1 checks if the top-ranked disease matches the ground truth. Additionally, we compute the CAR to measure whether the model prioritizes the top-$k$ diseases and avoids the bottom-$q$ diseases recommended by the upstream multimodal ML model.

To account for variations in phrasing or formatting that retain the same semantic meaning, we employ the SequenceMatcher algorithm[76]. This technique calculates a similarity ratio between two strings, effectively identifying matches even when textual representations differ slightly. For instance, it can align abbreviations with full names (e.g., "CdLS" and "Cornelia de Lange Syndrome") or account for other minor variations in disease name representation (e.g., “Simpson-golabi-behmel Syndrome, type 1” and “Type 1 Simpson-golabi-behmel Syndrome”.

We set the similarity threshold to 0.6 for HFA to allow greater flexibility, as this stage focuses on detecting plausible outputs rather than assessing strict correctness. This step ensures minimizes false negatives during the plausibility check. Conversely, stricter thresholds of 0.8 are applied for Top-10, Top-1, and CAR, as these later stages require higher precision to avoid false positives and ensure accurate ranking and matching. This balance between flexibility and stringency ensures a robust evaluation framework.

***Data sources***

For rare disease prediction, we primarily utilize the GMDB database (v1.0.9)[50,52], a multimodal medical resource containing frontal facial images of patients along with corresponding textual metadata, including demographic information and clinical HPO terms. This database is accessible to researchers in the medical field but requires an application process to obtain access to the data.

For external validation, we employ two independent data sources beyond the primary GMDB database, the Phenopacket-derived clinical notes[60]. This dataset comprises 5980 high quality synthesized clinical notes created by converting standardized phenotype profiles (called “Phenopackets”) into natural language narratives. These notes simulate realistic clinical documentation by transforming structured phenotypic data into flowing text with the linguistic variability commonly observed in medical records. We divide this dataset into two subsets: one containing set of diseases that overlap with our GMDB training set (1638 samples representing 72 diseases) and another with diseases that are completely disjoint from our training set. (4342 samples representing 396 diseases)

For tissue type classification, we employ the PanNuke database, a comprehensive pan-cancer dataset designed for nuclei instance segmentation and classification. This dataset encompasses 19 distinct tissue types, annotated through a semi-automated process and rigorously quality-checked by clinical pathologists. As a result, it offers a dataset whose characteristics closely mirror those encountered in clinical settings, with minimal selection bias.

***Supervised Fine-tuning (SFT) of LLMs***

**SFT Data Construction:** Although this is a multimodality dataset, for the purpose of fine-tuning LLMs, we only use the texts component of GMDB. We work with the samples that have non-null present features, or equivalently, at least one HPO id in the “present features” column of the metadata. In this case, we will do the following two steps: (1) Transform the HPO id(s) into real text data via the standard HPO dictionary and then concatenate them with a comma "," in between. For instance, the "HP:0000486; HP:0001263; HP:0010864" will become "Strabismus, Global developmental delay, Intellectual disability, severe." (2) Add patients' demographic information in the front. The image metadata of GMDB database contains patients' sex, age, and ethnicity information, which will be combined for LLM fine-tuning. For tissue type classification, we format our fine-tuning data in the simple Visual Question Answering (VQA) format. **Table S4** contains the sample SFT data used for both rare disease prediction and tissue type classification.

**LoRA fine-tuning of LLMs**: For fine-tuning, we utilize a single node with 4 NVIDIA A100 GPUs (40GB) and employ Low-Rank Adaptation (LoRA)[77] to reduce memory overhead. Our LoRA configuration is set with 16-bit precision, and the parameters are tailored to model size: for Llama 3.1-8B-Instruct and Gemma-2-9B-Instruct, we set the rank to 128 and scaling factor (lora alpha) to 64; for Llama 3.2-1B-Instruct, Llama 3.2-3B-Instruct and Gemma-2-2B-Instruct, we increase the rank to 256 and scaling factor to 128. A consistent LoRA dropout rate of 0.05 is set across all models to mitigate overfitting. For tissue type classification task, we set the rank to 128 and scaling factor to 64 for Llama3.2-Vision-11B-Instruct model.

### *Inference via Retrieval Augmented Generation (RAG)*

Retrieval-Augmented Generation (RAG) combines retrieval systems with LLMs to enhance the relevance and contextual accuracy of generated text, making it particularly well-suited for domain-specific tasks such as disease prediction. In this work, we utilize LlamaIndex[78] to construct our RAG database, as it offers a flexible interface for connecting both structured and unstructured medical knowledge sources to LLMs, enabling efficient and accurate retrieval. The database is built from two key biomedical resources: OMIM[79] and HPO[80]**.** To improve retrieval precision, we preprocess these datasets by systematically mapping HPO and OMIM terms to their corresponding disease names, ensuring that phenotypic and disease-related terminology is accurately aligned.

To further enhance the embedding representations for diseases and phenotypes, we adopt NeuML/pubmedbert-base-embeddings[81], a biomedical sentence transformer specifically trained on large-scale biomedical literature. This model captures the complex relationships and semantics of medical terms, allowing our system to understand the nuances of disease and phenotype descriptions more effectively. Together, these components enable our RAG pipeline to retrieve domain-specific, high-quality information, which is then used by the LLM to generate accurate and contextually appropriate outputs tailored to the disease prediction scenario.

### *Preference Optimization Frameworks*

**Overview:** MINT is a framework for aligning Large Language Models with multimodal biomedical expertise through preference optimization. In this study, we explore two implementations of the MINT framework: MINT with DPO (Direct Preference Optimization) and MINT with ORPO (Odds Ratio Preference Optimization). While traditional Reinforcement Learning from Human Feedback (RLHF) approaches like PPO[82-84] require reward model training, extensive hyperparameter tuning, and complex sampling procedures, DPO offers a more direct approach to preference alignment. The process begins with a base model, followed by SFT (Supervised Fine-tuning) to align the model's responses with specific requirements. The final step introduces a direct optimization objective that leverages the SFT model as a reference point to optimize the preference alignment between chosen and rejected responses. This approach enhances the model's domain-specific skills by learning from a dataset of human preferences without the complexity of reward modeling and policy optimization typically associated with RLHF methods.

ORPO, on the other hand, integrates preference alignment directly into the SFT phase in a single step. Its objective function leverages the odds ratio, offering a stable and effective approach to distinguish between preferred and non-preferred responses. This method can be applied effectively across various model sizes and demonstrates superior performance.

**Preference Learning Dataset Construction:** Due to the substantial time and resources required for manual data labeling, we employ the previously developed GestaltMML and PLIP models to build a dataset based on preference learning.

In each instance within the GMDB or PanNuke database, GestaltMML or PLIP produces a ranked list of potential labels (either disease names or tissue types). From this list, the top k labels (for example, k = 5, 10) are identified as preferred ("chosen") outputs, indicative of probable diagnoses, while the bottom q labels (for example, q = 10, 50) are classified as non-preferred ("rejected") outputs, suggesting fewer probable diagnoses. Generally, q is set larger than k to reflect that potential diagnoses are fewer than improbable ones. If the most probable prediction (Top-1 prediction) from GestaltMML or PLIP does not align with the actual prediction, it is substituted with the true prediction to ensure its presence at the top of the "chosen" outputs.

Thus, each case is structured into a pair of prompts: the input data sample is followed by the "chosen" and "rejected" labels, with the accurate prediction consistently presented first in the list. The sample of preference learning dataset are illustrated in **Table S4**.

**Prompting Strategies for Inference:** To optimize MINT for rare disease prediction and tissue type classification, we implement a dual-phase prompt strategy to prevent model collapse and catastrophic forgetting. This approach allows the model to strengthen diagnostic accuracy while ensuring outputs that are clear and clinically useful.

During the inference stage, we introduce a more detailed prompt format to ensure the model's outputs are presented in a consistent and structured way, making them easier to interpret for clinicians and data analysts. This phase's prompt requires a numbered list of diseases, with precise formatting guidelines that are easy to follow and interpret. The concrete prompts for inference are presented in **Table S4**.

The added structure in the inference prompt help MINT produce outputs that are not only diagnostically accurate but also neatly formatted and suitable for clinical review. By implementing this dual-phase prompt strategy, we ensure that MINT could generate flexible and insightful responses during training while providing reliable, structured outputs in inference, which is the key for supporting human interpretation in clinical workflows.

**Direct Preference Optimization (DPO):** DPO is a preference-based learning framework built upon a SFT model, designed to align model predictions with task-specific preferences. DPO builds directly on the SFT model by leveraging preference signals to prioritize chosen diagnoses $y_\omega$ over rejected diagnoses $y_l$, ensuring the model aligns its outputs to reflect user-defined preferences. To achieve this, DPO compares the output probabilities of the trainable actor model $P_\theta(\cdot)$ with those of a frozen reference model $P_{ref}(\cdot)$, which is a static copy of the SFT model. The reference model ensures that the preference-aligned actor model does not deviate too far from the base SFT model, thus preventing issues such as hallucinations or overly biased outputs. The DPO objective function is expressed as follows:

$$L_{DPO}\left(P_\theta; P_{ref}\right) = -\boldsymbol{E}_{(x,y_\omega,y_l)\sim D}\left[log\ \sigma\left(\beta \cdot log\ \frac{P_\theta(y_\omega|x)}{P_{ref}(y_\omega|x)} - \beta \cdot log\ \frac{P_\theta(y_l|x)}{P_{ref}(y_l|x)}\right)\right]$$

Here, $P_\theta(y|x)$ represents the conditional probability of the actor model, $P_{ref}(y|x)$ represents the frozen reference model, $x$ is the input phenotypic description, and $\beta$ is a scaling factor to control the distance between actor model and reference model. By maximizing the relative log likelihood of chosen response over rejected response while referencing the base SFT model, DPO ensures the effective alignment with preferences without compromising the foundational consistency of the model.

**MINT:** <u>Our default implementation, MINT with ORPO,</u> utilizes the Odds Ratio Preference Optimization (ORPO). Compared to DPO, the ORPO is a framework that combines Supervised Fine-Tuning (SFT) and preference learning in one stage.

Given a set of paired diagnoses $(y_\omega, y_l)$ for each patient case, ORPO minimizes the preference loss $L_{ORPO}$ which encourages higher scores for chosen diagnoses:

$$L_{ORPO} = E_{(x,y_\omega,y_l)}[L_{SFT} + \beta \cdot L_{OR}],$$

where $L_{SFT}$ represents the conventional causal language model negative log-likelihood (NLL) loss function to maximize the likelihood of generating the reference tokens. $L_{OR}$ maximizes the odds ratio between the likelihood of generating the favored response $y_{\omega}$ over disfavored response $y_l$. More specifically, the odds of $y$ given $x$ is defined by

$$Odds_{\theta}(x) := \frac{P_{\theta}(x)}{1-P_{\theta}(x)} .$$

Following the above, the odds ratio of a chosen response $y_{\omega}$ over $y_l$ , denote by $OR_{\theta}(y_{\omega}, y_l)$, is defined by

$$OR_{\theta}(y_{\omega}, y_l) \coloneqq \frac{Odds_{\theta}(x)}{Odds_{\theta}(x)},$$

and the $L_{OR} = -\log \sigma[\log OR_{\theta}(y_{\omega}, y_l)]$ . In other words, the log odds ratio was wrapped with the log sigmoid function so that $L_{OR}$ could be minimized by increasing the log odds ratio between $y_{\omega}$ and $y_l$.

Finally, in $L_{ORPO}$, the $L_{OR}$ is weighted with $\beta$ adapts the pretrained model to the domain-specific task and meanwhile enforces separation between chosen and rejected outputs (see **Figure S2**), compelling the model to push disfavored generations further down.

***Why no clipping in DPO/ORPO (vs. PPO-style RLHF)***

In PPO-style RLHF, a clipping mechanism is commonly used to stabilize policy-gradient updates by constraining the change in the policy ratio between successive iterations. In contrast, DPO/ORPO optimize the model directly from preference pairs using a supervised objective, rather than performing on-policy reinforcement learning with sampling-based policy updates. Concretely, DPO introduces a reference model and a temperature parameter that implicitly controls how far the optimized policy deviates from the reference, while ORPO combines token-level likelihood training with odds-ratio preference term in a single-stage objective. As a result, DPO/ORPO typically do not require PPO-style ratio clipping for stability. In practive, we still follow standard stabilization practices for neural training such as learning rate warm-up, conservative learning rates and optional gradient clipping. And we did not observe training instability attributable to preference optimization in our settings.

**Benchmark experiments**

For rare disease prediction, we perform five methodological comparisons on Llama 3.2-3B-Instruct: (1) direct inference using the base model, (2) base model with RAG support, (3) base model after supervised fine-tuning (SFT), (4) DPO, and (5) MINT. Additionally, we also assess other four relatively small open-source models of different sizes: Llama 3.2-1B-Instruct, Llama 3.1-8B-Instruct, Gemma-2-2B-Instruct, and Gemma-2-9B-Instruct. For larger models like Llama 3.1-70B-Instruct and Llama 3.1-405B-Instruct, we limit our evaluation to their direct inference capabilities using the base model, as implementing additional techniques is not feasible due to computational resource constraints faced by the research community. All the results besides Llama 3.2-3B-Instruct are presented in **Table S1**.

Beyond these eight experiments, we also evaluate other LLMs, including the closed-source GPT-4o (access time Nov. 2024), on one-shot learning using CoT prompts. Sample responses are included in the Supplementary Material.

For tissue type classification, we perform four methodological comparisons on Llama 3.2-Vision-11B-Instruct: (1) direct inference using the base model, (2) base model after supervised fine-tuning (SFT), (3) DPO, and (4) MINT.

## DISCUSSION

In this study, we introduce MINT (Multimodal Integrated kNowledge Transfer), a novel and effective framework for aligning unimodal LLMs with multimodal biomedical data through preference optimization, with ORPO serving as the default backbone mechanism in our implementation. Our comprehensive experiments demonstrate MINT's effectiveness in two complex biomedical tasks: rare disease prediction from clinical texts and tissue type classification from histological images. These results support MINT's potential as a versatile approach for enhancing domain-specific capabilities of large language models (LLMs) without compromising their general language and visual understanding.

Despite MINT's promising results, several important limitations must be acknowledged. First, MINT's performance diminishes substantially in zero-shot scenarios with completely unseen disease classes, as evidenced by the lower accuracy on disjoint disease subsets in our external validation. Importantly, this degradation is consistent with the limited label exposure in the constructed preference dataset: among the 396 unique disjoint ground-truth diseases, 377 (95.2%) never appear in either the chosen or rejected candidate lists used for preference optimization (**Table S5**), providing little direct supervision signal for those unseen labels. While this is expected for any fine-tuning approach, it highlights the need for complementary techniques such as RAG when applying MINT in real-world clinical settings where extremely rare, previously unseen diseases may be encountered. Second, the quality of the upstream multimodal model significantly influences MINT's effectiveness. MINT inherits both the strengths and the weaknesses of its upstream model. If the upstream model contains biases or errors, these may propagate to MINT-enhanced LLM. This risk is particularly relevant in medical applications where demographic or geographic biases in training data could lead to disparities in diagnostic accuracy. Third, while MINT significantly improves specific task performance, it still falls short of specialized encoder-based classification models in some scenarios. This gap reflects the fundamental architectural differences between encoder models optimized for classification and decoder models designed for generative tasks. The auto-regressive nature of text generation introduces additional complexity compared to direct label selection, which may limit the upper bound of MINT's performance on classification tasks.

The core innovation of MINT lies in its unified approach to aligning unimodal models with multimodal expertise through preference optimization. Rather than using traditional knowledge transfer methods that require parallel training of teacher and student models, MINT uses multimodal models as preference data generators to create structured datasets of likely and unlikely predictions. These preference datasets then guide the alignment of downstream LLMs through ORPO, enabling them to benefit from patterns identified in multimodal data while preserving their general capabilities. This approach offers several distinct advantages: First, MINT effectively bridges the modality gap between multimodal machine learning models and text-only or image-only LLMs. Our rare disease prediction experiments demonstrate that MINT can align LLMs with patterns from GestaltMML, which utilizes facial images, demographic information and clinical phenotypes. Although the downstream LLMs consume only text, they still benefit from facial image data indirectly through the preference-learning dataset, which encapsulates inter-modal relationships learned by GestaltMML. Second, MINT demonstrates similarly impressive performance in the tissue type classification task with preference data from PLIP, helping models prioritize relevant tissue types and avoid implausible ones more effectively than baseline approaches. Third, MINT's preference optimization approach enables more nuanced learning than traditional supervised fine-tuning by incorporating both positive and negative examples during training, developing a more discriminative understanding of subtle features that differentiate similar conditions, as evidenced in our tissue type classification case study. Fourth, MINT-enhanced decoder-based LLMs show promising generalization capabilities compared to traditional encoder-only classifiers, particularly in zero-shot scenarios with previously unseen disease classes, demonstrating the decoder architecture's inherent flexibility and reasoning advantages. Finally, key design choices contribute to MINT's success: the balanced Acceptance-over-Rejection ratio provides clear contrastive learning signals; the odds ratio loss delivers stable gradients even with limited training data; and the approach preserves general model capabilities while enhancing domain-specific performance, making it particularly valuable for biomedical applications.

Looking forward, the MINT framework opens several promising directions for future research in biomedical AI. First, hybrid approaches combining MINT with retrieval mechanisms could address the zero-shot limitations while preserving MINT's strong in-distribution performance. Our external validation results on disjoint disease subsets, where RAG outperformed MINT, suggest that such hybrid approaches might be particularly valuable for rare disease prediction, where unseen presentations are common. Second, extending MINT to additional biomedical tasks beyond disease prediction and tissue type classification could demonstrate its broader utility. Potential applications include medical image analysis, genomic interpretation, drug discovery, and clinical decision support. Each domain would require domain-specific upstream models but could benefit from MINT's flexible knowledge transfer approach. Third, investigating the interpretability[85] of MINT-enhanced models represents an important clinical research direction. While MINT improves predictive performance, understanding the reasoning behind its predictions remains crucial for clinical trust and adoption. Future work could explore techniques for extracting and visualizing the diagnostic patterns learned through preference optimization. For tissue classification specifically, integrating region-of-interest highlighting could help pathologists understand which histological features most influenced the model's predictions. Finally, prospective clinical evaluations in real-world healthcare settings will be essential to validate MINT's practical utility. Such studies should assess not only diagnostic accuracy but also impact on clinical decision-making, time-to-prediction for rare conditions, and overall patient outcomes.

In conclusion, MINT represents an advancement in biomedical AI by bridging the gap between specialized multimodal models and general-purpose large language models. By leveraging preference optimization to transfer domain-specific knowledge, MINT enhances the capabilities of open-source LLMs in critical biomedical tasks while maintaining their inherent flexibility and reasoning capabilities. The framework demonstrated consistent effectiveness across multiple tasks, models, and external validation datasets suggests broad applicability in the biomedical domain. As LLMs continue to evolve as interfaces for healthcare information, approaches like MINT that can effectively integrate specialized domain knowledge while preserving general capabilities will be increasingly valuable for advancing precision medicine and improving patient care.

## ACKNOWLEDGEMENTS

We thank the GestaltMatcher Database (GMDB) which provides a collection of curated medical photography of genetic syndromes for training the multimodal model used in the current study. We thank patients and their families for contributing facial photos and phenotype descriptions to enable the establishment of computational models. This project is supported by NIH grant HG013031, OD037960 and the CHOP Research Institute.

## DECLARATIONS

### *Availability of data and materials*

The GestaltMatcher Database (GMDB, v1.0.9) analyzed in this study is publicly available via the GestaltMatcher data portal (https://db.gestaltmatcher.org). All source code, environment specifications, and the complete computational workflow (including SLURM batch scripts), are openly released under the MIT License at https://github.com/WGLab/MINT-LLM. This study did not generate any new biological specimens, cell lines, or other physical materials.

### *Competing interests*

The authors declare ***NO*** competing interests.

### *Use of Generative AI*

The authors used generative LLMs only for proofreading, checking grammar, and correcting typos to improve the readability of the paper.

## Figures

**Figure 1. Overview of the MINT framework for transferring multimodal knowledge to Large Language Models.** The framework consists of several pipelines**: (1) Upstream Pipeline**: A multimodal classifier integrates test and image modality input data to generate top-k most-likely and bottom-q least-likely predictions, which are organized into chosen (preferred) and rejected (non-preferred) responses in natural language to form a preference dataset**. (2) Downstream Pipeline-SFT**: Standard supervised fine-tuning of base language or vision-language. (**3) Downstream Pipeline-MINT with DPO**: Direct Preference Optimization approach that uses a frozen reference model (initialized from SFT) and a trainable policy model, optimizing with KL-divergence and maximum likelihood objectives. **(4) Downstream Pipeline-MINT with ORPO (default)**: Our proposed unified framework combining negative log likelihood and odds ratio loss in a single step, directly optimizing the relative probabilities between chosen and rejected responses without requiring a separate reference model.

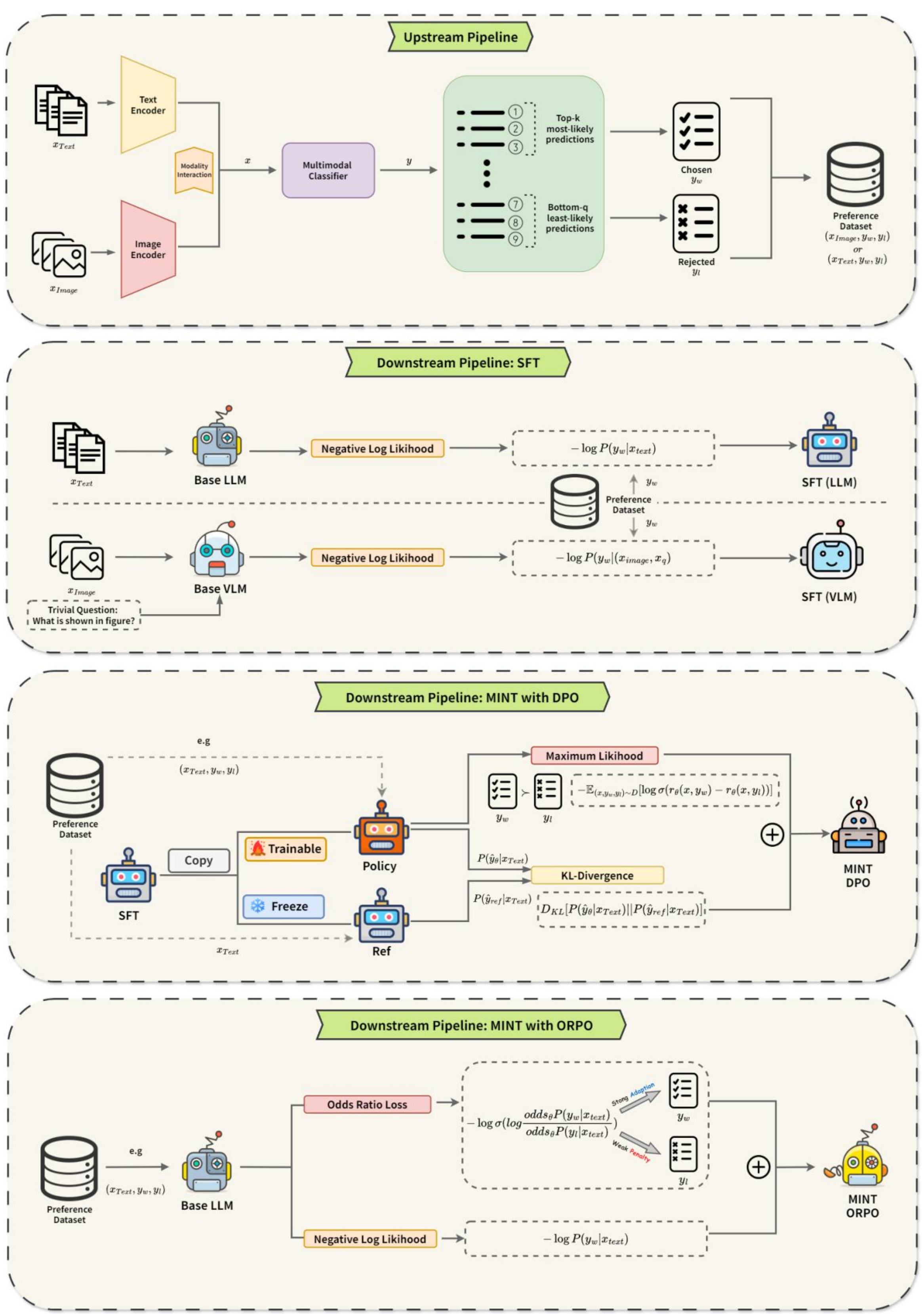

**Figure 2. Performance evaluation of rare disease prediction techniques using Llama 3.2-3B-Instruct model**. **(a)** Comparison of model performance across four evaluation metrics: Hallucination-Free Accuracy (HFA), Top-10 accuracy, Top-1 accuracy, and Coverage-Avoidance Ratio (CAR) for five approaches: Base Model, RAG, SFT, MINT with DPO, and MINT with ORPO (color from light to dark respectively).**(b)** Effect of varying Acceptance-over-Rejection (AoR) ratios on MINT performance, showing optimal performance at balanced ratio (AoR=1). **(c)** Radar chart comparing performance on six language understanding benchmarks, demonstrating preserved general capabilities across all fine-tuning techniques.

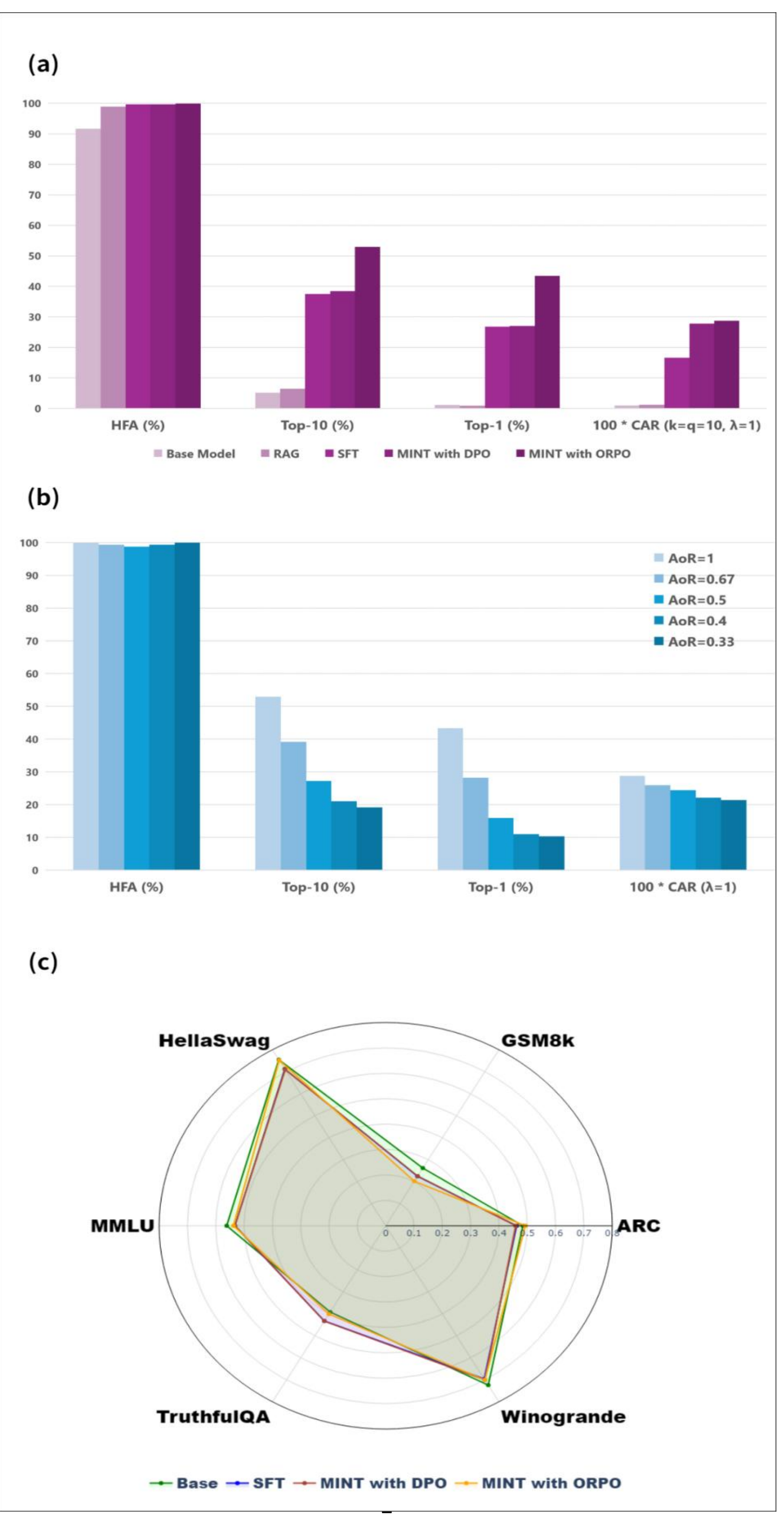

**Figure 3. Performance on tissue type classification using different fine-tuning techniques on the Llama 3.2-Vision-11B-Instruct foundation model. (a)** Bar Chart showing performance metrics across four evaluation criteria: Hallucination Free Accuracy (%), Top-5 Accuracy (%), Top-1 Accuracy (%), and CAR (Coverage-Avoidance Ratio). Four fine-tuning approaches are compared: Base Model, SFT, MINT with DPO, MINT with ORPO (color from light to dark respectively). **(b)** Radar chart comparing performance across multiple general vision-language capabilities for different fine-tuning techniques.

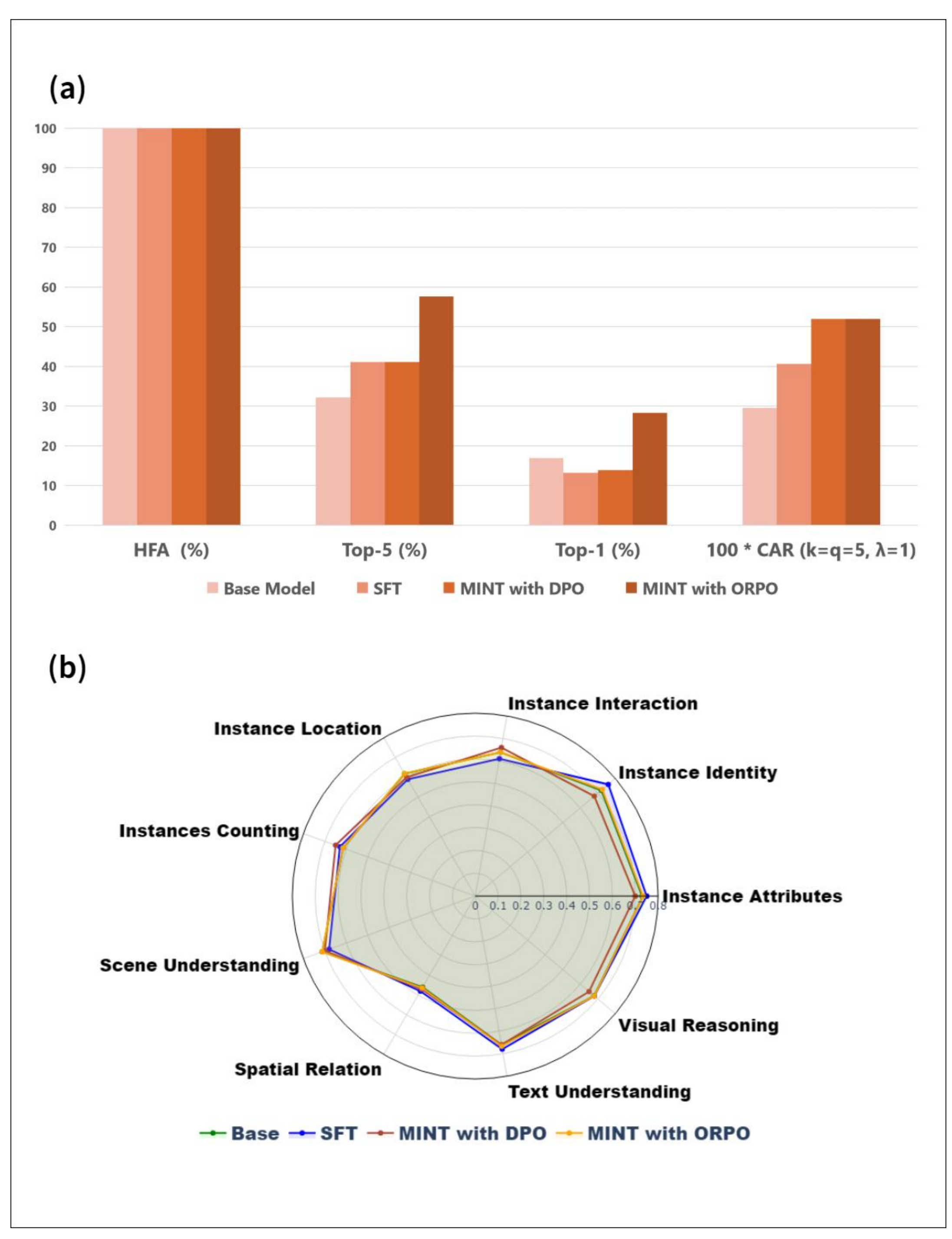

**Figure 4. Comparative analysis of tissue type classification performance between Base model, SFT, and MINT for similar-looking bile duct and colon tissues.** The figure demonstrates how MINT improves discrimination between histologically similar tissues by leveraging both positive and negative training examples. Top panel shows bile duct tissue classification: a representative training sample with corresponding chosen and rejected tissue types (left), and four testing samples with their respective ranks assigned by Base, SFT, and MINT (Right). Bottom panel shows the same analysis for colon tissue classification. Green values represent ranking for the ground truth tissue class, while red values indication rankings for the visually similar confused class. Lower values represent higher confidence (rank 1 is the highest). Average ranks across all test samples are shown in both panels. Average ranks across all test samples are shown at the bottom of each panel. **'MINT' refers to our default implementation of the MINT framework using ORPO.**

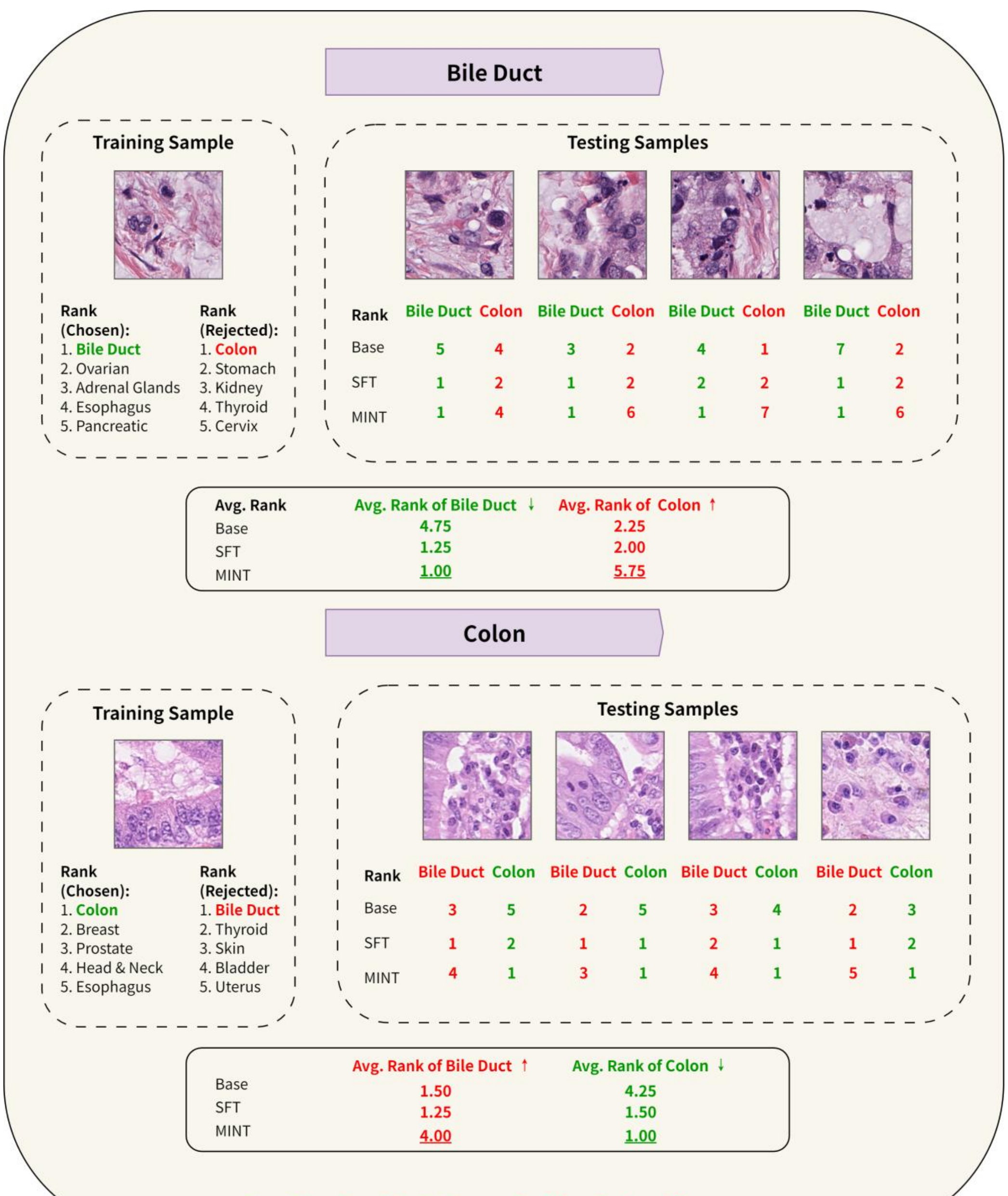
Bile Duct
Training Sample
Rank (Chosen):
1. Bile Duct
2. Ovarian
3. Adrenal Glands
4. Esophagus
5. Pancreatic
Rank (Rejected):
1. Colon
2. Stomach
3. Kidney
4. Thyroid
5. Cervix
Testing Samples
Rank
Bile Duct Colon Bile Duct Colon Bile Duct Colon Bile Duct Colon
Base 5 4 3 2 4 1 7 2
SFT 1 2 1 2 2 2 1 2
MINT 1 4 1 6 1 7 1 6
Avg. Rank
Avg. Rank of Bile Duct ↓
Avg. Rank of Colon ↑
Base 4.75 2.25
SFT 1.25 2.00
MINT 1.00 5.75
Colon
Training Sample
Rank (Chosen):
1. Colon
2. Breast
3. Prostate
4. Head & Neck
5. Esophagus
Rank (Rejected):
1. Bile Duct
2. Thyroid
3. Skin
4. Bladder
5. Uterus
Testing Samples
Rank
Bile Duct Colon Bile Duct Colon Bile Duct Colon Bile Duct Colon
Base 3 5 2 5 3 4 2 3
SFT 1 2 1 1 2 1 1 2
MINT 4 1 3 1 4 1 5 1
Avg. Rank of Bile Duct ↑
Avg. Rank of Colon ↓
Base 1.50 4.25
SFT 1.25 1.50
MINT 4.00 1.00
Green Class: Ground Truth Class
Red Class: Confused Class

## Tables

**Table 1. Summary of datasets, upstream multimodal ML model and downstream LLM for the two biomedical tasks evaluated in this study.**

| Detailed Components of MINT Framework | | |
|---|---|---|
| Task | Rare Disease Prediction | Tissue Type Classification |
| Dataset | GestaltMatcher Database | PanNuke Database |
| Images | Frontal Facial Image | Nucleus Image |
| Texts | Age, Sex, Ethnicity, HPO terms | Tissue type |
| Upstream Multimodal ML Model | GestaltMML | Pathology Language and Image Pre-Training (PLIP) |
| Downstream LLM | Llama 3.2-3B-Instruct | Llama 3.2-Vision-11B-Instruct |
| # of Training Sample | 6522 | 5436 |
| # of Testing Sample | 386 | 2330 |
| # of Labels (Categories) | 528 | 19 |

**Table 2. Evaluation of Llama 3.2-3B-Instruct variants on Phenopacket-derived clinical notes.** Results show three evaluation metrics: Hallucination Free Accuracy (%), Top-10 Accuracy, and Top-1 Accuracy for five model variants based on the Llama 3.2-3B-Instruct architecture. Each row represents a different approach: Base Model, RAG, SFT, DPO and MINT. Results are presented for three data configurations: overlapping diseases (1638 samples with 72 diseases shared with training data), disjoint diseases (4342 samples with 396 diseases unseen in training data), and the combined dataset (5980 samples). Bold values indicate best performance for each metric in each configuration. **'MINT' refers to our default implementation of the MINT framework using ORPO.**

| ***Performance on Phenopacket-derived Clinical Notes (N=1638) (72 overlapping diseases)*** | | | |
|---|---|---|---|
| Model (Llama 3.2-3B-Instruct) | HFA (%) | Top-10 Accuracy (%) | Top-1 Accuracy (%) |
| Base | 99.94 | 5.13 | 3.42 |
| RAG | **100.00** | 30.11 | 26.45 |
| SFT | 99.94 | 46.40 | 14.65 |
| DPO | 99.94 | 60.99 | 26.56 |
| MINT | 99.89 | **66.91** | **47.56** |
| ***Performance on Phenopacket-derived Clinical Notes (N=4342) (396 disjoint diseases)*** | | | |
| Model (Llama 3.2-3B-Instruct) | HFA (%) | Top-10 Accuracy (%) | Top-1 Accuracy (%) |
| Base | 99.99 | 20.00 | 11.29 |
| RAG | **100.00** | **24.17** | **13.80** |
| SFT | 100.00 | 9.16 | 8.46 |
| DPO | 99.99 | 9.48 | 7.71 |
| MINT | 99.90 | 10.48 | 7.00 |
| ***Performance on Phenopacket-derived Clinical Notes (N=5980)*** | | | |
| Model (Llama 3.2-3B-Instruct) | HFA (%) | Top-10 Accuracy (%) | Top-1 Accuracy (%) |
| Base | 99.98 | 15.58 | 8.98 |
| RAG | **100.00** | 25.96 | 17.62 |
| SFT | 99.98 | 20.25 | 10.18 |
| DPO | 99.98 | 24.24 | 13.47 |
| MINT | 99.90 | **27.05** | **20.23** |

## Supplementary Materials

**Figure S1 (A). Rare Disease Prediction Case Study (Text-only).** This example demonstrates how MINT resolves diagnostic ambiguity by leveraging domain-specific preference signals. The input describes a patient with *Cornelia de Lange Syndrome (CdLS)*, presenting with generic phenotypes (e.g., "Intellectual disability", "Hirsutism") alongside a specific facial feature ("Synophrys"). The **SFT baseline (Left)** is misled by the high overlap of generic symptoms (specifically Hirsutism) and incorrectly predicts the confuser class *Hypertrichosis* as Rank 1. In contrast, the **MINT model (Right)**, aligned with preferences derived from a facial-analysis upstream model, correctly identifies *CdLS* as Rank 1. This suggests MINT implicitly learns to weight specific discriminative features (like Synophrys) higher than generic overlapping symptoms to rule out "hard negative" confusers. **(B). Tissue Type Classification Case Study (Vision-only).** This example illustrates MINT's improved discriminative capability on histologically similar tissues. The input image shows **Colon** tissue, characterized by glandular epithelial structures that are morphologically similar to **Bile Duct** tissue. The **SFT baseline (Left)** fails to distinguish these subtle differences, incorrectly prioritizing the confuser class *Bile Duct* (Rank 1). The **MINT model (Right)**, which was trained on preference pairs where *Bile Duct* was explicitly rejected against *Colon*, successfully resolves this ambiguity. It correctly predicts *Colon* as Rank 1 and significantly demotes the visually similar confuser *Bile Duct* to Rank 5, demonstrating that preference optimization sharpens the decision boundary between structurally related categories.

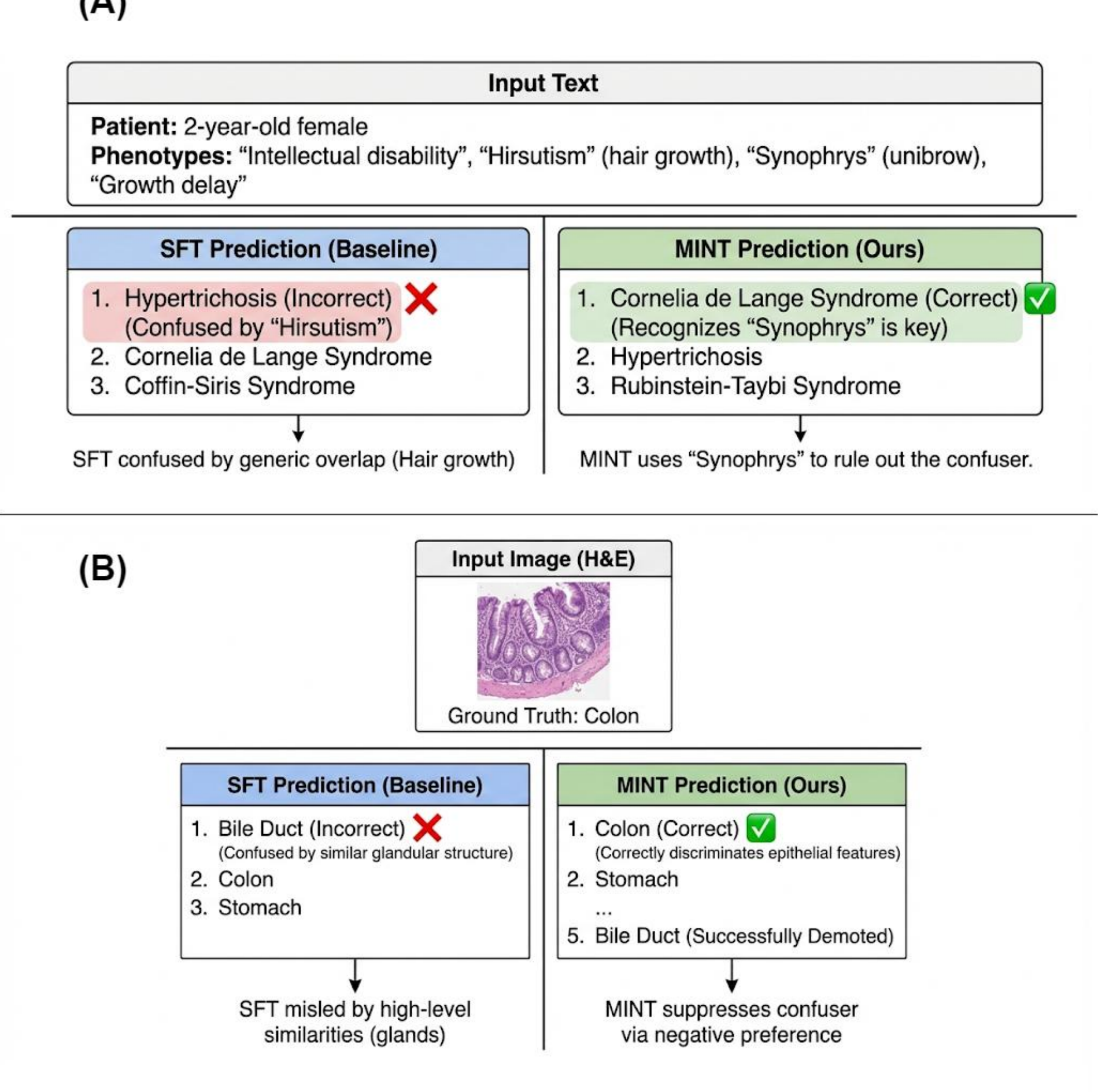

**Table S1. Main Results on Performance of various Open-Source LLMs on Rare Genetic Diseases prediction.** The mean and standard deviation are computed using three different random seeds. **'MINT' refers to our default implementation of the MINT framework using ORPO.**

| *Performance of LLMs on Rare Genetic Diseases Prediction with Various Add-on Techniques* | | | | | |
|---|---|---|---|---|---|
| **LLM Enhancement Pipeline (LLMEP)** | | **Evaluation Metrics (EM)** | | | |
| Base Model | Add-on Technique | HFA (%) | Top-10 (%) | Top-1 (%) | $CAR_{(k=10,q=10,\lambda=1)}$ |
| ***Llama 3.2-1B-Instruct*** | N/A | 95.04±1.15 | 4.24±0.23 | 0.65±0.18 | 0.0292±0.0016 |
| | +RAG | 76.68±4.74 | 2.21±0.17 | 0.94±0.22 | 0.0271±0.0003 |
| | +SFT | 88.82±3.60 | 8.30±0.01 | 1.08±0.05 | 0.0255±0.0019 |
| | +DPO | 89.22±0.37 | 7.90±0.18 | 1.63±0.26 | 0.1785±0.0054 |
| | **+MINT** | **99.48±0.37** | **34.28±1.21** | **23.91±3.41** | **0.2120±0.0018** |
| ***Llama 3.2-3B-Instruct*** | N/A | 91.61±0.59 | 5.19±0.01 | 1.06±0.02 | 0.0522±0.0025 |
| | +RAG | 98.90±0.28 | 6.52±0.05 | 0.91±0.18 | 0.0517±0.0049 |
| | +SFT | 99.71±0.42 | 37.53±0.68 | 26.86±0.12 | 0.1665±0.0051 |
| | +DPO | 99.63±0.21 | 38.49±0.58 | 27.07±0.19 | 0.2780±0.0016 |
| | **+MINT** | **99.91±0.15** | **52.99±1.69** | **43.49±0.57** | **0.2877±0.0035** |
| ***Llama 3.1-8B-Instruct*** | N/A | 95.07±1.45 | 6.63±0.21 | 1.67±0.17 | 0.0797±0.0005 |
| | +RAG | 99.61±0.55 | 10.95±0.10 | 1.92±0.16 | 0.0776±0.0006 |
| | +SFT | 87.57±3.91 | 13.86±1.27 | 6.52±0.37 | 0.1415±0.0009 |
| | +DPO | 84.93±0.43 | 14.02±0.19 | 5.94±0.56 | 0.1618±0.0100 |
| | **+MINT** | **97.28±2.02** | **25.91±3.29** | **16.97±0.92** | **0.2326±0.0009** |
| ***Gemma-2-2B-Instruct*** | N/A | 100.00±0.00 | 6.99±0.00 | 1.79±0.03 | 0.1065±0.0063 |
| | +RAG | 98.07±1.36 | 9.84±0.23 | 2.01±0.03 | 0.1100±0.0033 |
| | +SFT | 89.14±1.27 | 7.60±0.25 | 5.13±0.08 | 0.1259±0.0017 |
| | +DPO | 85.70±3.41 | 7.48±0.08 | 5.07±0.00 | 0.1737±0.0061 |
| | **+MINT** | **99.04±1.36** | **16.32±0.00** | **8.04±0.01** | **0.1743±0.0051** |
| ***Gemma-2-9B-Instruct*** | N/A | 99.56±0.63 | 12.37±0.82 | 5.03±0.16 | 0.1368±0.0047 |
| | +RAG | 96.71±0.01 | 10.10±0.00 | 2.99±0.00 | 0.1104±0.0001 |
| | +SFT | 91.98±0.01 | 14.12±0.04 | 5.29±0.02 | 0.1331±0.0011 |
| | +DPO | 92.07±1.68 | 13.91±0.19 | 4.88±0.07 | 0.2315±0.0023 |
| | **+MINT** | **99.61±0.55** | **33.80±1.25** | **23.40±0.85** | **0.2465±0.0122** |
| ***Llama3.1-70B-Instruct*** | N/A | 99.74±0.16 | 21.24±0.27 | 11.92±0.08 | 0.1122±0.0031 |
| ***Llama3.1-405B-Instruct*** | N/A | 100±0.00 | 26.17±0.22 | 11.66±0.08 | 0.1221±0.0019 |
| ***Llama3.3-70B-Instruct*** | N/A | 100±0.00 | 26.94±0.21 | 11.66±0.08 | 0.1221±0.0019 |

**Table S2: Performance of MINT on Llama3.2-3B-Instruct for Rare Disease Prediction Across Different Acceptance-over-Rejection (AoR) ratios $k/q$.** Recall that $k$ denotes the number of positive labels, while $q$ represents the number of negative samples within our preference-based dataset generated using GestaltMML. In this experiment, $k$ is consistently set to 10 across all ratio variations.

| *Performance of MINT on Llama3.2-3B-Instruct Across Various Acceptance-over-Rejection (AoR) Ratios $k/q$* | | | | |
|---|---|---|---|---|
| $k/q$ | HFA | Top-10 | Top-1 | CAR ($k = 10, q = 10 * k/q$) |
| 10/10 | **100.00** | **53.88** | **44.04** | **1.63** |
| 10/15 | 99.48 | 38.13 | 28.29 | 1.47 |
| 10/20 | 98.96 | 26.06 | 15.96 | 1.38 |
| 10/25 | 99.97 | 21.07 | 10.99 | 1.25 |
| 10/30 | 100.00 | 20.21 | 10.36 | 1.21 |

**Table S3. Main Results on Performance of Llama 3.2-11B-Vision-Instruct on Tissue Type Classification.** The mean and standard deviation are computed using three different random seeds. **'MINT' refers to our default implementation of the MINT framework using ORPO.**

| ***Performance of LLMs on Tissue Type Classification with Various Add-on Techniques*** | | | | |
|---|---|---|---|---|
| Method/Metric | HFA (%) | Top-5 (%) | Top-1 (%) | $CAR_{(k=5,q=5,\lambda=1)}$ |
| Llama 3.2-11B-Vision-Instruct | 100.00 | 32.21±0.0784 | 16.96±0.0900 | 0.2965±0.0068 |
| Llama 3.2-11B-Vision-Instruct +SFT | 100.00 | 41.46±0.0900 | 13.22±0.1700 | 0.4065±0.0037 |
| Llama 3.2-11B-Vision-Instruct +DPO | 100.00 | 41.46±0.1700 | 13.91±0.0850 | 0.5196±0.0079 |
| Llama 3.2-3B-Instruct +MINT | 100.00 | **57.58±0.195** | **28.41±0.0850** | **0.5203±0.0043** |

**Table S4. Overview of Prompts Utilized in This Study.** For SFT and MINT, which involve supervised training, labels (responses) are also included. For inference, only the prompts are presented. **'MINT' refers to our default implementation of the MINT framework using ORPO.**

| Overview of Prompts Utilized In This Study | | |
|---|---|---|
| Techniques | Medical Task | Sample Prompt Content |
| **Direct Inference** | Rare Disease prediction | **System Message:** *"You are a genetic counselor. Your task is to identify potential rare diseases based on given phenotypes. Follow the output format precisely."*<br><br>**User Message:** *"A 1 year(s) 0 month(s) old female patient of Caucasian descent exhibits the following phenotypes: long face global developmental delay overgrowth, soft, doughy skin abnormality of the cardiovascular system, abnormality of the kidney scoliosis. What are the potential diagnoses?*<br><br>*Based on this information, provide a numbered list of EXACTLY 10 potential rare diseases.*<br><br>*Use EXACTLY this format:*<br><br>*POTENTIAL_DISEASES:*<br><br>*\n1. 'Disease1'\n2. 'Disease2'\n3. 'Disease3'\n4. 'Disease4'\n5. 'Disease5'\n6. 'Disease6'\n7. 'Disease7'\n8. 'Disease8'\n9. 'Disease9'\n10. 'Disease10'*<br><br>*Ensure all disease names are in single quotes, and there are exactly 10 in the list. Do not deviate from this format or add any explanations."* |
| **Direct Inference** | Tissue Type Classification | **System Message:**<br>*"You are a professional pathologist analyzing nucleus images. For each image, identify the most likely tissue types based on the nuclear morphology. Follow the output format precisely."*<br>**User Message:**<br>*Based on the morphological features observed in the provided nucleus image, which tissue types could this nucleus originate from?:*<br>*Use EXACTLY this format:*<br><br>*MOST_LIKELY_TISSUE:*<br>*\n1. 'tissue type 1'\n2. 'tissue type 2'\n3. 'tissue type 3'\n4. 'tissue type 4'\n5. 'tissue type 5'*<br><br>*Ensure all tissue types are in single quotes, and List exactly 5 tissue types in decreasing order of likelihood. Do not deviate from this format or add any explanations.* |
| **SFT** | Rare Disease prediction | **System Message:** *"You are a genetic disease diagnosis assistant."*<br><br>**User Message:** *"A 1 year(s) 0 month(s) old female patient of Caucasian descent exhibits the following phenotypes: long face global developmental delay overgrowth, soft, doughy skin abnormality of the cardiovascular system, abnormality of the kidney scoliosis. What are the potential diagnoses?"*<br><br>**Response:** *POTENTIAL_DISEASES:\n1. hypoplastic left heart syndrome\n2. cutis laxa\n3. geroderma osteodysplasticum; go\n4. phelan-mcdermid syndrome; phmds\n5. simpson-golabi-behmel syndrome, type 1; sgbs1\n6. kbg syndrome;* |

| | | *kbgs\n7. ogden syndrome; ogdns\n8. neurodevelopmental disorder with neonatal respiratory insufficiency, hypotonia, and feeding difficulties; nedrihf\n9. mucolipidosis ii alpha/beta\n10. williams-beuren syndrome; wbs* |
|---|---|---|
| **SFT** | Tissue Type Classification | **User Message:** ***"****Based on the morphological features observed in the provided nucleus image, which tissue types could this nucleus originate from?"*<br>**Input Image:**<br>**Response:** *'MOST_LIKELY_TISSUE:\n\n1. Breast\n2. Colon\n3. Head & Neck\n4. Testis\n5. Prostate'* |
| **MINT** | Rare Disease prediction | **System message**: *You are a genetic disease diagnosis assistant*<br>**Prompt:**<br>*A 3 years and 6 months old male patient of European descent exhibits the following phenotypes: long palpebral fissure, growth delay, decreased body weight, eversion of lateral third of lower eyelids, feeding difficulties. What are the most likely diagnoses?*<br>**Chosen Response:**<br>*POTENTIAL_DISEASES:\n1. hypoplastic left heart syndrome\n2. cutis laxa\n3. geroderma osteodysplasticum; go\n4. phelan-mcdermid syndrome; phmds\n5. simpson-golabi-behmel syndrome, type 1; sgbs1\n6. kbg syndrome; kbgs\n7. ogden syndrome; ogdns\n8. neurodevelopmental disorder with neonatal respiratory insufficiency, hypotonia, and feeding difficulties; nedrihf\n9. mucolipidosis ii alpha/beta\n10. williams-beuren syndrome; wbs*<br>**Rejected Response**<br>*POTENTIAL_DISEASES:\n1. alzahrani-kuwahara syndrome; alkus\n2. microcephalic osteodysplastic primordial dwarfism, type ii; mopd2\n3. neurodevelopmental disorder with structural brain anomalies and dysmorphic facies; nedbaf\n4. cerebral creatine deficiency syndrome\n5. cardiac, facial, and digital anomalies with developmental delay; cafdadd\n6. clark-baraitser syndrome; clabars\n7. neurodevelopmental disorder with hypotonia and cerebellar atrophy, with or without seizures; nedhcas\n8. intellectual developmental disorder, autosomal recessive 5; mrt5\n9. robinow syndrome\n10. developmental and epileptic encephalopathy 76; dee76'* |
| **MINT** | Tissue Type Classification | **User Message:** ***"****Based on the morphological features observed in the provided nucleus image, which tissue types could this nucleus originate from?"*<br>**Input Image:**<br>**Chosen Response:** *'MOST_LIKELY_TISSUE:\n\n1. Breast\n2. Colon\n3. Head & Neck\n4. Testis\n5. Prostate'*<br>**Rejected Response:** *'MOST_LIKELY_TISSUE:\n\n1. Stomach\n2. Thyroid\n3. Skin\n4. Cervix\n5. Kidney'* |

**Table S5. Disjoint disease label space analysis.** We compare the 396 unique ground-truth diseases in the Phenopacket disjoint split against the diseases appearing in the preference candidates extracted from the upstream construction; 377/396 (95.2%) never appear in either the chosen or rejected lists, and 19/396 (4.8%) appear in both (chosen-only = 0; rejected-only = 0).

| Split | #Unique GT diseases | % in chosen | % in rejected | % in both | % never |
|---|---|---|---|---|---|
| Disjoint | 396 | 4.8% (19/396)* | 4.8% (19/396)* | 4.8% (19/396) | 95.2% (377/396) |

**Table S6. Sensitivity Analysis of MINT Performance on Rare Disease Prediction and Tissue Type Classification with Varying Preference Dataset Sizes.** We evaluated the downstream models aligned via MINT using varying percentages (20% to 100%) of the total available preference dataset. For Rare Disease Prediction (Top Panel), we used Llama 3.2-3B-Instruct with Top-10 accuracy. For Tissue Type Classification (Bottom Panel), we used Llama 3.2-Vision-11B-Instruct with Top-5 accuracy. The mean and standard deviation are computed using three different random seeds.

| *Rare Disease Diangosis (Llama 3.2-3B-Instruct)* | | | | | |
|---|---|---|---|---|---|
| **Preference Dataset Size (%)** | **Number of Training Pairs** | **HFA (%)** | **Top-10 Accuracy (%)** | **Top-1 Accuracy (%)** | **CAR (k=10, q=10)** |
| 20% | 1,304 | 98.45 ± 0.52 | 19.32 ± 1.15 | 11.45 ± 0.85 | 0.0950 ± 0.0042 |
| 40% | 2,608 | 99.12 ± 0.35 | 34.60 ± 1.05 | 24.30 ± 0.92 | 0.1840 ± 0.0035 |
| 60% | 3,913 | 99.55 ± 0.20 | 43.20 ± 0.88 | 32.50 ± 0.75 | 0.2310 ± 0.0028 |
| 80% | 5,217 | 99.80 ± 0.15 | 49.75 ± 0.65 | 39.10 ± 0.45 | 0.2650 ± 0.0019 |
| **100%** | **6,522** | **99.91 ± 0.15** | **52.99 ± 1.69** | **43.49 ± 0.57** | **0.2877 ± 0.0035** |
| *Tissue Type Classification (Llama 3.2-Vision-11B)* | | | | | |
| **Preference Dataset Size (%)** | **Number of Training Pairs** | **HFA (%)** | **Top-5 Accuracy (%)** | **Top-1 Accuracy (%)** | **CAR (k=10, q=10)** |
| 20% | 1,087 | 100.00 ± 0.00 | 38.45 ± 0.55 | 19.12 ± 0.42 | 0.3350 ± 0.0051 |
| 40% | 2,174 | 100.00 ± 0.00 | 45.20 ± 0.62 | 22.35 ± 0.50 | 0.4120 ± 0.0045 |
| 60% | 3,262 | 100.00 ± 0.00 | 50.15 ± 0.45 | 25.10 ± 0.35 | 0.4680 ± 0.0032 |
| 80% | 4,349 | 100.00 ± 0.00 | 54.80 ± 0.30 | 27.20 ± 0.25 | 0.4950 ± 0.0028 |
| **100% (Full)** | **5,436** | **100.00 ± 0.00** | **57.58 ± 0.20** | **28.41 ± 0.09** | **0.5203 ± 0.0043** |

**Table S7. Performance Comparison with Domain-Specific Specialized Model (MedGemma-1.5-4b-it) on GMDB In-Distribution Test Set.** We compared the MINT-enhanced Llama-3.2-3B model against a specialized medical baseline of similar size, MedGemma-1.5-4b-it (4B parameters). The evaluation was performed on the in-distribution GMDB test set. While the specialized model outperforms the generalist base model, MINT enables the smaller generalist model to achieve superior diagnostic accuracy on the target domain.

| Model | Type | Parameters | HFA (%) | Top-10 Accuracy (%) | Top-1 Accuracy (%) |
|---|---|---|---|---|---|
| Llama-3.2-3B-Instruct (Base) | Generalist | 3B | 91.61 | 5.19 | 1.06 |
| MedGemma-1.5-4b-it | Specialized (Medical) | 4B | 99.20 | 32.45 | 19.80 |
| Llama-3.2-3B-Instruct + MINT | Generalist + MINT | 3B | **99.91** | **52.99** | **43.49** |